\documentstyle[psfig,graphics]{mn2e}

\title [Kinematics, ages and metallicities]
{Kinematics, ages and metallicities for F and G type stars
in the solar neighbourhood}

\author[Karata\c{s} et al.]
       {Y.~Karata\c{s},$^1 \thanks{E-mail: karatas@istanbul.edu.tr}$
       S.~Bilir$^1$ and W.J. Schuster$^{2}$\\
  $^1$Istanbul University Science Faculty, Department of Astronomy and Space 
       Sciences, 34119 University-Istanbul, Turkey\\
    $^2$Observatorio Astron\'omico Nacional, UNAM, Apartado
        Postal 877, C.P. 22800, Ensenada, B.C., M\'exico\\}

\date{Accepted 2005 month day.
      Received year month day;
      }

\pagerange{\pageref{firstpage}--\pageref{lastpage}}
\pubyear{2005}

\begin{document}

\maketitle

\label{firstpage}

\begin{abstract}

A new metallicity distribution and an age-metallicity relation are presented
for 437 nearby F and G turn-off and sub-giant stars selected from radial velocity 
data of Nidever et al.  Photometric metallicities are derived from $uvby-H\beta$ 
photometry, and the stellar ages from the isochrones of Bergbusch \& VandenBerg 
as transformed to $uvby$ photometry using the methods of Clem et al.

The $X$ (stellar-population) criterion of Schuster et al., which combines
both kinematic and metallicity information, provides 22 thick-disk stars.
$\sigma_{\rm W} = 32 \pm 5$ km s$^{-1}$, $<V_{\rm rot}> = 154 \pm 6$ km s$^{-1}$
and $<[M/H]> = -0.55 \pm 0.03$ dex  for these thick-disk stars, which is in
agreement with values from previous studies of the thick disk.  $\alpha$-element
abundances which are available for some of these thick-disk stars show the typical
$\alpha$-element signatures of the thick disk, supporting the classification
procedure based on the $X$ criteria.

Both the scatter in metallicity at a given age and the presence of old, metal-rich
stars in the age-metallicity relation make it difficult to decide whether or
not an age-metallicity relation exists for the older thin-disk stars.  For
ages greater than 3 Gyr, our results agree with the other recent studies that there
is almost no correlation between age and metallicity, $\Delta ([M/Fe])/\Delta$(age)$
= -0.01 \pm 0.005$ dex Gyr$^{-1}$. For the 22 thick-disk stars there is a range in ages of
7--8 Gyr, but again almost no correlation between age and metallicity.

For the subset of main-sequence stars with extra-solar planets the age-metallicity
relation is very similar to that of the total sample, very flat, the main
difference being that these stars are mostly metal-rich, $[M/H] \ga -0.2$ dex.
However, two of these stars have $[M/H] \sim -0.6$ dex and have been classified as
thick-disk stars.  As for the total sample, the range in ages for these stars with
extra-solar planetary systems is considerable with a nearly uniform distribution over 
$3 \la$ age $\la 13$ Gyr.

\end{abstract}


\section{Introduction}
Metallicity, kinematic, and age data of F and G type stars are of fundamental
importance to study the metallicity distribution and the age-metallicity 
relation of the Galactic disk. These F and G stars are unique objects
for studies of the history of the Galactic disk in the sense
that photometric systems, such as the Str\"omgren-Crawford $uvby-H\beta$,
can be used to derive many of their physical parameters, such as the
interstellar reddening, metallicity, and age.  Also, the combined information 
of stellar photometric abundances and kinematics can be used to separate
stellar populations and to summarize the main properties of a given population.

With this aim, based on the accurate radial velocity observations of
Nidever et al.\ (2002), a data set has been created which includes parallaxes,
proper motions, and $uvby-H\beta$ photometry of late-F, G and K stars.
Accurate parallaxes and proper motions have become available for large numbers of
these stars from the $Hipparcos$ and $Tycho$ catalogues (ESA\ 1997) and $Tycho$--2 
(Hog et al.\ 2000) catalogue.
With this data set, which is large compared to the sample of
Edvardsson et al.~(1993), but small compared to the
samples of Feltzing, Holmberg \& Hurley (2001) and Nordstr\"om et al.~(2004),
the metallicity distribution and age-metallicity relation in the solar vicinity
are determined.
Much effort has been devoted to the determination of reliable individual
isochrone ages for our sample of stars. For this task, the isochrones of
Bergbusch \& VandenBerg (2001) and VandenBerg (2003)
as transformed to the $uvby$ system by the empirically constrained
colour-temperature relations of Clem et al.~(2004), are adopted.

The sample of Nidever et al.\ (2002) contains 889 stars from the Doppler planet
search project at the Keck and Lick Observatories (Butler et al.~2000,
Vogt et al.~2000). This sample contains mostly main-sequence and sub-giant 
stars from F7 to M5, within 50 pc. Chromospherically active stars have been
removed from this list as well as stars with known stellar companions within 2 $\arcsec$ 
(including known spectroscopic binaries). This list provides several advantages
for our study of the age-metallicity relation: (a) it includes many cooler 
dwarf stars and so avoids one of the selection affects discussed by
Feltzing et al. \ (2001); when these cool dwarfs are left out of a sample, 
the old, metal-rich stars are preferentially excluded depleting the upper 
right-hand corner of an age-metallicity diagram. (b) the radial velocities 
are especially accurate (systematic errors $\la \pm 0.3$ km s$^{-1}$) 
and precise (random errors $\la \pm0.1$ km s$^{-1}$) 
(Nidever et al.\ 2002) providing us with especially good kinematic data 
for these stars allowing us to separate more cleanly the stellar populations 
within the sample, such as the separation between the thin- and thick-disk
stars. (c) As suggested by Gim\'enez (2000) the $uvby$ photometry can be
used to refine the survey lists of planetary-search projects and to study the
metallicities and ages of stars found to have planets.

This paper is organized as follows:  Section~2 describes the data; in Section~3
the calculation procedure for the space velocities is given; the derivation
of astrophysical parameters:  reddening, metallicity, absolute magnitudes, and
ages is given in Section~4, plus age-metallicity diagrams and a discussion of these; 
and finally the conclusions in Section~5.

\section{Data}

\subsection{Photometry}

For the 889 stars in the catalogue of Nidever et al.\ (2002),
$uvby$ photometry of 437 stars and $H\beta$ photometry for 214 
have been collected from the web site 
of the General Catalogue of Photometric 
Data\footnote[1]{http:// obswww.unige.ch/gcpd/gcpd.html} of Hauck \& Mermilliod (1998).
This $uvby-H\beta$ photometry ($b-y$, $m_{\rm 1}$, $c_{\rm 1}$,  $H\beta$) of these 
437/214 stars are given in Cols.~2--5 of Table 1, respectively. In Table 1 this 
$uvby$ photometry of the 437 F and G main-sequence, turn-off and sub-giant stars 
covers the ranges $0.087 \leq m_{\rm 1} \leq 0.497$, $0.116 \leq c_{\rm 1} \leq 0.492$,
$0.313 \leq (b-y) \leq 0.580$.  The 214 stars with $H\beta$ values fall in the
range, $2.530 \leq H\beta \leq 2.680$. Visual magnitudes for our sample stars, 
all brighter than $V=10$ mag, and $(B-V)$ colours, given in Cols.~7--8 of Table 1,
respectively, were taken from the $Hipparcos$ and $Tycho$ Reference Catalogues 
(ESA\ 1997).

\begin{table*}
\caption{Stellar parameters for our sample stars. The columns are described in the text.
The full version of this table is available online at 
http://www.blackwellpublishing.com/products/journals/MNR/MNR?/
mnr?sm.htm}
{\scriptsize
\begin{tabular}{lrrrrrcrrr}
\hline
Name & $b-y$ &   $m_{1}$ &    $c_{1}$ &  $H_{\beta}$ & $E(b-y)$ &   
$V$ &  $(B-V)$ &  $M_{V}$ &      $[M/H]$ \\
\hline
 HD    377 &0.391&0.204&0.322&    - & 0.004 &7.59&0.626&4.55&-0.02\\
 HD    400 &0.332&0.147&0.386&2.615 &-0.014 &6.21&0.524&3.61&-0.35\\
 HD   1388 &0.382&0.186&0.346&    - & 0.001 &6.51&0.599&4.41&-0.08\\

\hline
\end{tabular}
} 
\end{table*}

\subsection{Radial Velocities, Parallaxes and Proper Motions}

The barycentric radial velocities $(V_{\rm rad})$, with a typical accuracy of $\la 0.3$
km s$^{-1}$, were taken from Nidever et al.\ (2002).
Parallaxes ($\pi$), the proper motion components ($\mu_{\alpha}cos\delta$, $\mu_{\delta}$)
and their associated uncertainties were taken mainly from the $Hipparcos$
and $Tycho$ catalogues (ESA\ 1997) and  $Tycho$--2 catalogue (Hog et al.~2000).
Parallaxes and their uncertainties for three stars (HD 185295, HD 192020 and HD 1854) 
not found in the $Hipparcos$ and $Tycho$ catalogues were taken from the catalogue of Kharchenko (2001).
The radial velocities $(V_{\rm rad})$, parallaxes ($\pi$), proper motion
components ($\mu_{\alpha}cos\delta$, $\mu_{\delta}$) and their associated
uncertainties are listed in Table 2. Most of the $Hipparcos$ parallaxes in our sample 
have relative errors that are much less than 20 per cent. In Tables 1--3 those stars
with extra-solar planetary systems have a ``P'' after their HD number in Col.~1; these
stars are discussed further in Section~4.4 and especially in Fig.~9c.

\begin{table*}
\caption{Hipparcos astrometric, radial velocity and kinematic data of
our sample stars. The columns are described in the text. 
The full version of this table is available online at 
http://www.blackwellpublishing.com/products/journals/MNR/MNR?/
mnr?sm.htm}
{\scriptsize
\begin{tabular}{lrrrrrrr}
\hline
Star Name &$\mu_{\alpha}cos\delta$~~~~& $\mu_{\delta}~~~~$ & $\pi$ ~~~~~~~& 
$V_{rad}$~~~~& $U$~~~~ & $V$~~~~ & $W$~~~~~~\\
&(mas yr$^{-1}$)& (mas yr$^{-1}$)& (mas)~~~~~~& (km s$^{-1})$ & (km s$^{-1})$ & (km s$^{-1})$ & (km s$^{-1})$\\
\hline

 HD  377 &  87.26$\pm$0.93&  -0.90$\pm$0.53&25.15$\pm$0.97&  1.18&-14.41& -7.07&-3.79\\
 HD  400 &-114.72$\pm$0.53&-124.60$\pm$0.43&30.26$\pm$0.69&-15.14& 27.64&-10.36&-7.85\\
 HD 1388 & 396.97$\pm$1.09&  -0.60$\pm$0.53&38.24$\pm$0.85& 28.50&-42.81&-16.06&-33.82\\
 
\hline

\end{tabular}
}  
\end{table*}

\subsection{Comparison of our sample with other samples}

Our sample was selected based on the existence of accurate radial velocities
for 889 FGKM main-sequence and sub-giant stars of Nidever et al.\ (2002).
The sample used in this work is reduced to only 437 stars, for which $uvby-H\beta$
photometry is available, to be able to determine a photometric metallicity for each star.
Our sample is free from binary contamination according to Nidever et al.\ (2002).
The accurate radial velocities of Nidever et al.\ (2002) are suitable for the detection
of extra-solar planets, and so the high accuracy of their radial velocities offers us an 
opportunity to derive very good kinematic parameters and thus the age-metallicity relations
for sub-classes of F and G stars.

Compared with the samples of Edvardsson et al.~(1993) and Chen et al.~(2003),
the size of our sample is large, whereas compared to the samples and metallicity distributions 
of Feltzing et al. \ (2001) and Nordstr\"om et al.~(2004),
the size of our sample is small but covering a wide metallicity range, $-2.0 <$[M/H]$ <+0.5$ dex,
with a very high percentage of the stars having $-1.0 <$[M/H]$ < +0.5$ dex.
Our sample is not affected by one of the main biases discussed by Feltzing et al. \ (2001) 
in the sense that old, metal-rich stars are amply represented (see Fig.~9 below);
the presence of these old, metal-rich stars produces an age-metallicity relation very similar
to other modern age-metallicity relations, such as those mentioned above.

One advantage of having accurate radial velocities for our sample together with the 
$\pm0.12-0.14$ dex uncertainty in [M/H] (Schuster \& Nissen 1989, and see Section~4.3
below) is that it allows us to determine well the $X$ cut-offs of the stellar populations
from $V_{\rm rot}$ and [M/H], where $X$ is the stellar-population criterion defined in
Schuster, Parrao \& Contreras-Mart\'{\i}nez (1993), a linear combination of $V_{rot}$ and [M/H].

\section{Galactic Space Velocities}

Galactic space velocity components ($U$, $V$, $W$) were computed
by applying the algorithms and the transformation matrices of
Johnson \& Soderblom (1987) to the basic data: celestial coordinates
($\alpha$, $\delta$), proper motion components ($\mu_{\alpha}cos\delta$,
$\mu_{\delta}$), the radial velocity $(V_{\rm rad})$ and the parallax ($\pi$) of
each star in Table 2, where the epoch of J2000 was
adopted as described in the International Celestial Reference System (ICRS)
of the $Hipparcos$ and $Tycho$ Catalogues. The transformation 
matrices use the notation of a right-handed system. Therefore, $U$, $V$, $W$
are the components of a velocity vector of a star with respect to the Sun, 
where $U$ is positive toward the Galactic center
($l = 0^{o}, b = 0^{o}$); $V$ is positive in the direction of the Galactic
rotation ($l=90^{o}, b=0^{o}$); and $W$ is positive toward the north
Galactic pole ($b=90^{o}$). $U$, $V$, $W$ velocities are heliocentric. 
The velocity components $U', V', W'$ have been corrected for the local
solar motion $(U, V, W)_{\odot} = (+10.0, +14.9, +7.7)$ km s$^{-1}$ with respect 
to the local standard of rest (LSR, Allen \& Santillan 1991)

\begin{figure}
\resizebox{8cm}{10.5cm}{\includegraphics*{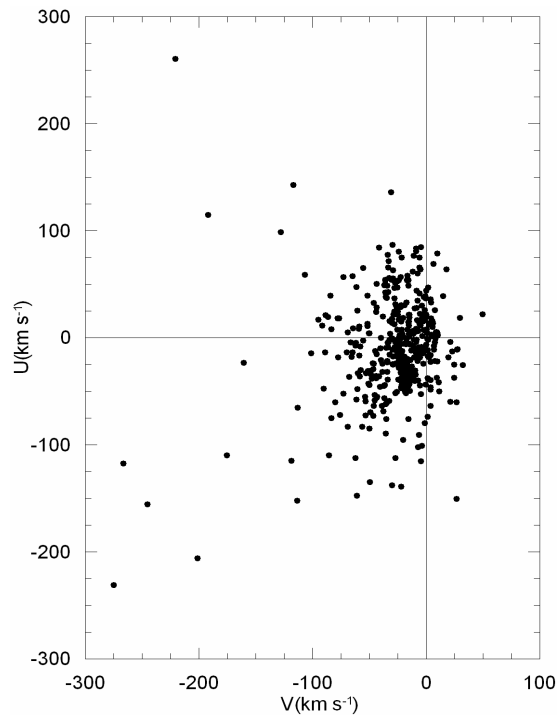}}
\caption{The Bottlinger diagram for our sample. The velocities are heliocentric.}
\end{figure}

The distribution of these Galactic heliocentric velocities on the $(U, V)$ plane
is displayed in Fig.~1. This Bottlinger diagram
shows a strong concentration around $|U| \la 75$ km s$^{-1}$  and
$-75  \la V \la$ 50 km s$^{-1}$, where disk-like components dominate.
However, Fig.~1 also shows more halo-like components having
$|U| \ga$ 100 km s$^{-1}$  and  $V \la -100$ km s$^{-1}$.
Fig.~1 shows that there exist no extreme retrograde velocities in our sample;
only three stars are retrograde, and these all have $V \geq -300$ km s$^{-1}$
(Compare to Fig.~3 of Schuster et al. \ 1993).

\section{Derivation of astrophysical parameters}

\begin{figure}
\resizebox{8cm}{8cm}{\includegraphics*{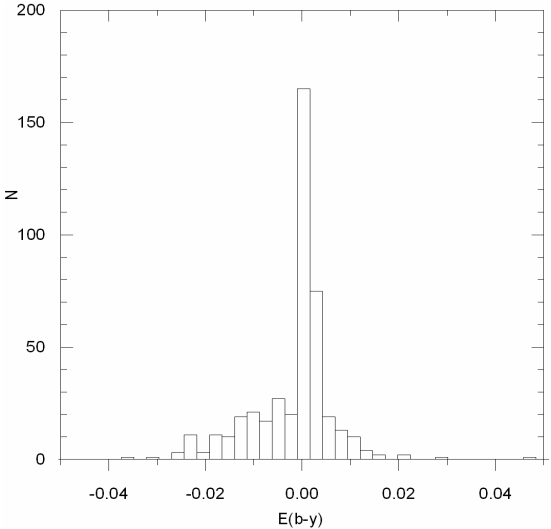}}
\caption{$E(b-y)$ histogram of all stars in our sample.}
\end{figure}

\subsection{Reddening}

The interstellar extinction for our sample stars has been measured due to its effect 
on the derivation of the metallicities and absolute magnitudes. $E(b-y)$ of 214
stars with $2.530 \leq H\beta \leq 2.680$ were estimated from an intrinsic
colour calibration of $(b-y)_{o}$ vs $H\beta$ (Schuster \& Nissen 1989). 
A few iterations were made to obtain $E(b-y)$, with a consistency of 
$\pm0.0001$ for $(b-y)_{o}$.  For 223 stars which lack $H\beta$, $E(B-V)_{\rm S}$ 
was obtained from the Schlegel, Finkbeiner \& Davis (1998) reddening maps. We adopt a $V$ 
band absorption $A_{\rm V}=3.1 E(B-V)$ and assume that the dust layer has a scale height
$H = 125$ pc. Thus, the reddening $E(B-V)_{\rm S}$ for a given star is reduced
compared to the total reddening  $E(B-V)_{\rm T}$ by a factor $(1-e^{-d sin(b)/H})$, 
where {\it b} and {\it d} are the Galactic latitude and distance, respectively.
Note, however, that Arce \& Goodman (1999) caution that Schlegel et al. \ (1998) reddening maps 
overestimate the reddening values when the colour excess $E(B-V)_{\rm S}$ is more than 
0.15 mag. Hence, we have adopted (Schuster et al.\ 2004) a slight revision of 
the Schlegel et al. \ (1998) reddening estimates, via an equation, 
$E(B-V)_{\rm A} = 0.10 + 0.65(E(B-V)_{\rm S}-0.10)$ when $E(B-V)_{\rm S} > 0.10$, 
otherwise $E(B-V)_{\rm A} = E(B-V)_{\rm S}$, where $E(B-V)_{\rm A}$ indicates the adopted 
reddening estimate.  
Figure 3 of Schuster et al.\ (2004) shows that there is no significant systematic 
difference between the reddenings obtained from these two methods, the intrinsic colour
calibration and the reddening maps. The final adopted $E(B-V)_{S}$ values of 223 
stars are transformed into $E(b-y)$ via the equation $E(B-V)=1.35E(b-y)$ given
by Crawford \ (1975).

Fig.~2 shows the $E(b-y)$ distribution of the 437 stars. As expected, the stars
with a median distance of 36 pc show a strong peak at about $E(b-y)=0.000$, with
somewhat asymmetric wings, the negative tail being slightly bigger than the positive,
displaced only 0.005 mag from symmetry.  Reddenings less than 0.015 are mostly
due to the observational errors of the photometry (Nissen \ 1994). For our final sample,
reddening corrections have been applied to only six stars, those with $E(b-y) \geq 0.015$;
for all the remaining stars no reddening correction has been applied.  For these six stars,
the indices $c_{1}$ and $m_{1}$ are dereddened by the relations, $c_{o}=c_{\rm 1}-0.2E(b-y)$ 
and $m_{o}=m_{\rm 1} + 0.3E(b-y)$, given by Crawford \ (1975).  The final reddenings,
$E(b-y)$, are given in Col.~6 of Table 1.

\subsection{Absolute magnitudes}

The absolute visual magnitudes, $M_{\rm V}$, are needed to determine the ages of stars in our
sample from the $M_{\rm V}$, $(b-y)$ plane. Although relative errors ($\sigma_{\pi}/\pi$) 
of $Hipparcos$ parallaxes of the stars in Table 2 are less than 0.20, the Lutz-Kelker 
correction (Lutz \& Kelker 1973) has been applied to the absolute magnitudes of our 
sample using the following relation constructed from their data:

\begin{eqnarray}
\delta M_{\rm V}= a(\sigma_{\pi}/\pi)^{3}+
b(\sigma_{\pi}/\pi)^{2}+c(\sigma_{\pi}/\pi)+d
\end{eqnarray}
where $a=-63.0303$, $b=-1.0736$, $c=-0.3286$, $d=0.0008$.

The relation $M_{\rm V}=V_{o} - 5log (1/\pi) + 5 + \delta M_{\rm V}$ 
allows us to determine absolute magnitudes when $Hipparcos$
parallaxes of the stars in our sample are available. 
Here, $V_{o}$ magnitude is corrected for reddening using $V_{o} = V - A_{\rm V}$.
The absolute magnitudes of our sample are presented in Col.~9 of Table 1.

\subsection{The metallicity distribution and the separation of population types}

In order to estimate the metallicities of our sample stars from the $uvby-H\beta$
photometry presented in Table 1, the metallicity ([Fe/H]) calibrations of
Schuster \& Nissen (1989) have been used. We assume here that $[M/H] = [Fe/H]$.
However, Twarog, Anthony-Twarog \& Tanner (2002) report that the metallicity calibration of G stars 
given by Schuster \& Nissen (1989) has a systematic error that is strongly colour
dependent, reducing the estimated [M/H] for stars above $[M/H] \sim -0.2$ dex for the 
very reddest stars, $(b-y)_{\rm 0} \ga 0.47$. Martell \& Laughlin (2002) give an 
improved $uvby-H\beta$ metallicity relation for F, G and early K stars. In order 
to see this systematic effect, in Fig.~3 the differences between the metallicities 
derived from Martell \& Laughlin (2002) minus those from Schuster \& Nissen (1989) 
have been plotted versus those derived from the Schuster \& Nissen (1989) formulae. 
In Fig.~3 it is worth noting that only the stars with $(b-y)_{\rm 0}\ga 0.47$ and
$[M/H] \ga -0.03$ dex, show a significant difference, increasing to a maximum at $[Fe/H]
\approx +0.07$ dex and then decreasing again.  So, for the final metallicities of this 
paper, the [Fe/H] calibration given by Martell \& Laughlin (2002) has been applied 
to only the stars with $(b-y)_{\rm 0} > 0.47$ and $[M/H] \geq -0.2$ dex; for the
remaining stars, the metallicities estimated via the formulae of 
Schuster \& Nissen (1989) are adopted. The metallicities of our stars are given 
in the Col.~10 of Table 1.

\begin{figure}
\resizebox{8cm}{4cm}{\includegraphics*{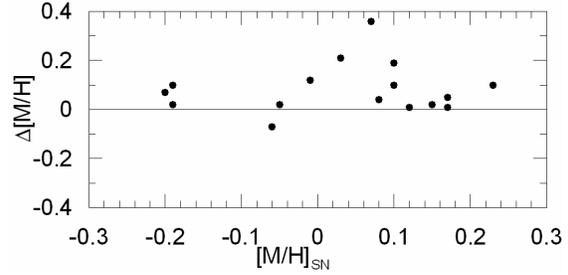}}
\caption{Differences of the metallicities derived from the formulae of Martell 
\& Laughlin (2002) minus those from Schuster \& Nissen (1989) versus the 
values from Schuster \& Nissen (1989) $([M/H]_{\rm SN})$, for the stars with
$(b-y)_{\rm 0} > 0.47$ in our sample.  Note that the systematic errors become 
obvious only for $[M/H]_{\rm SN}$ greater than about solar, maximize at 
$[M/H]_{\rm SN} \approx +0.07$ dex, and then become small again at $[M/H]_{\rm SN} 
\approx +0.15$ dex.}
\end{figure}

\begin{figure}
\resizebox{8cm}{8cm}{\includegraphics*{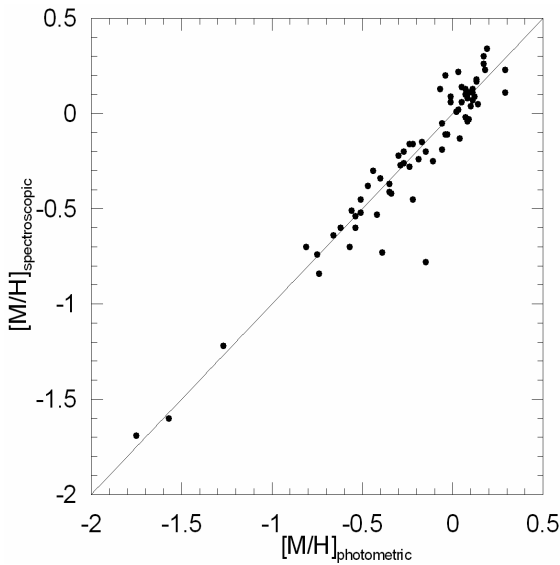}}
\caption{$uvby-H\beta$ photometric metallicities of 71 stars versus the 
spectroscopic ones taken from the Cayrel de Strobel et al. \ (2001) 
catalogue.  The one-to-one relation is drawn as a solid line.}
\end{figure}

In Fig.~4, for 71 stars, our photometrically derived [M/H] values are plotted
against the spectroscopic values of the Cayrel de Strobel, Soubrian \& Ralite (2001) 
catalogue.  A least square fit to the points of Fig.~4 yields a regression curve of
$[M/H]_{\rm spec}=0.99(\pm 0.04)[M/H]_{\rm phot} - 0.01 (\pm 0.02)$ for
the metallicity range $-2.0 <$[M/H]$<+0.5$ dex. The agreement is excellent with
a scatter of $\sigma = \pm 0.12$, excluding the large deviation of HD 202573, 
which is a giant star (Henry et al.~2000) and which is outside the range of the 
Schuster \& Nissen (1989) [Fe/H] calibrations.. It is clear from Fig.~4 that
accurate metal abundances can be derived from $uvby-H\beta$ photometry, using 
the combined calibrations as described above.

For the empirical [Fe/H] calibrations of Schuster \& Nissen (1989) the 
estimated standard deviations of a single photometric determination of [Fe/H]  were
$\pm 0.14$ at [Fe/H] $\approx -0.5$ dex, and $\pm 0.21$ at [Fe/H] $\approx -1.5$ dex.
Comparisons by Feltzing et al. \ (2001) of photometric abundances from 
the Schuster \& Nissen  (1989) calibration equations with abundances from two recent 
spectroscopic studies Edvardsson et al.~(1993) and Chen et al.~(2000)
have shown that these error estimates are overly conservative; they find a
scatter of only $\pm 0.10-0.11$ for their more metal-rich group, in very good
agreement with the value of $\pm 0.12$ found above from a somewhat less homogeneous
spectroscopic data set.

\begin{figure}
\resizebox{8cm}{8cm}{\includegraphics*{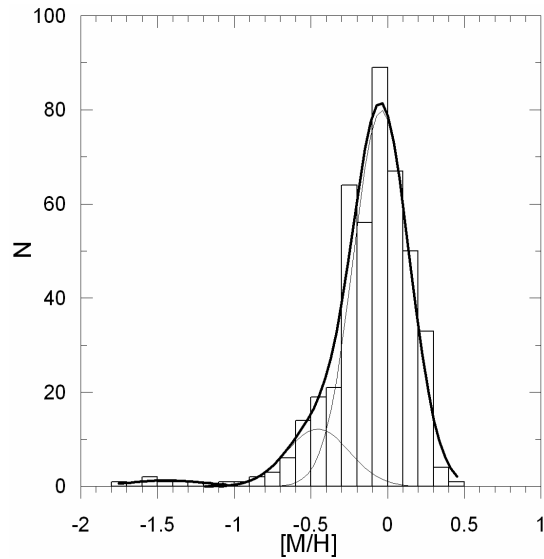}}
\caption{Distribution of the metallicities of our stars.
Curves are the fitted Gaussian distributions, which give the mean
metal abundances of $<[M/H]> = -0.04, -0.45$, and $-1.30$ dex for the
thin and thick disks, and halo (thin curves), respectively, and their sum (heavy curve).}
\end{figure}

In Fig.~5 the distribution of metallicity, [M/H], for our sample stars
is presented, and also shown is a fit to the histogram using three Gaussians.
The sample is mix of all the stellar populations that are represented in 
the solar neighbourhood, the thin and thick disks, and the halo.  As is expected, in 
the solar neighbourhood, the majority of the stars belong to the disk with the thin disk
dominating, a significant but smaller contribution from the thick disk, and only a few
halo stars ($\approx 8$; see Fig.~6 below).  In Fig.~5 two main components can be seen, one
with $-0.9 \la [M/H] \la +0.5$ dex (the disk contribution: thin plus thick) and
the other with $[M/H] \la -0.9$ dex (mostly the halo stars).  A rough Gaussian fit has
been made to this metallicity histogram using the mathematical package 
{\it`Origin'} \footnote[2]{http:// www.originlab.com}.
One disk-like component has $<[M/H]> \sim -0.04$ dex, a dispersion of $0.37$, and
is nearly symmetric. Our sample contains an obvious contribution from the thick disk
with a Gaussian having a mean abundance of $<[M/H]> \sim -0.45$ dex and a dispersion
of $0.41$. There are eight halo stars with $<[M/H]> \la -0.9$ dex (see also Fig.~6).
Nevertheless, thin and thick-disk stars are not clearly distinguishable from Fig.~5.
There exists no clean, straightforward procedure for separating the stars of these
different disk populations, which overlap considerably in the metallicity distribution.
Kinematics are usually invoked to improve the thin/thick disk separation, and more recently 
[$\alpha/Fe$] abundances (see for example, Bensby, Feltzing \& Lundstr\"om \ 2003).

\begin{table*}
\caption{Kinematic and abundance data of thick-disk stars.
The parameter $X$ defined by Schuster et al. \ (1993) in Col.~3 denotes
the linear combination of $V_{\rm rot}$ and $[M/H]_{\rm phot}$.
$[Fe/H]_{\rm spec}$, $\alpha$-elements and $[\alpha/Fe]$ are given in Cols.~6-10.
$[\alpha/Fe]$ is defined as (1/4)([Mg/Fe] + [Si/Fe] + [Ca/Fe] + [Ti/Fe]).
The references are as follows: 1, Bensby et al. (2005); 2, Pomp\'eia,  Barbuy \&  Grenon (2003); 
3, Cenarro et al. (2001); 4, Cayrel de Strobel et al. \ (2001); 5, Fulbright (2000); 
6, Mishenina et al. (2004); 7, Friel (1987); 8, Santos, Israelian \& Mayor (2004)}
\begin{tabular}{lcrccrrrrrr}
\hline
Name &  $V_{\rm rot}$ & $X$ &$[M/H]_{\rm phot}$ &$[Fe/H]_{\rm spec}$& [Mg/Fe]&  [Si/Fe] & [Ca/Fe]  & [Ti/Fe] & $[\alpha/Fe]$ & Refs.\\
      ~~&~~(km s$^{-1}$) &   &  &   &   &   &  &  & &\\
\hline
 HD   3795 &145 &-18 &-0.39 &-0.59 &       0.31 &       0.22 &       0.24 &       0.29 &       0.27 &          1 \\
 HD  16623 &143 &-16 &-0.50 &-0.60 &       0.25 &       0.10 &      -0.02 &       0.00 &       0.08 &          2 \\
 HD  19034 &166 &-20 &-0.41 &      &            &            &            &            &            &            \\
 HD  22879 &149 &-12 &-0.74 &-0.84 &       0.40 &       0.27 &       0.27 &       0.31 &       0.31 &          1 \\
 HD  29587 &185 &-20 &-0.59 &-0.61 &       0.31 &       0.17 &       0.17 &       0.19 &       0.21 &          1 \\
 HD  30649 &155 &-16 &-0.56 &-0.50 &            &            &       0.08 &            &            &          3 \\
 HD  32387 &173 &-19 &-0.55 &      &            &            &            &            &            &            \\
 HD  37213 &179 &-20 &-0.50 &      &            &            &            &            &            &            \\
 HD  63077 &174 &-14 &-0.81 &-0.83 &            &            &            &            &            &          4 \\
 HD  65583 &146 &-14 &-0.62 &-0.67 &       0.28 &       0.38 &       0.29 &       0.13 &       0.27 &          5 \\
 HD  68017 &174 &-21 &-0.42 &-0.42 &       0.31 &       0.21 &            &            &            &          6 \\
 HD  88725 &202 &-21 &-0.65 &-0.70 &       0.32 &       0.26 &       0.19 &       0.25 &       0.26 &          5 \\
 HD 101259 &128 &-19 &-0.23 &      &            &            &            &            &            &            \\
 HD 102158 &116 &-13 &-0.47 &-0.46 &            &            &            &            &            &          4 \\
 HD 104556 & 74 &-11 &-0.30 &-0.64 &            &            &            &            &            &          7 \\
 HD 111515 &151 &-15 &-0.58 &-0.52 &            &            &            &            &            &          4 \\
 HD 114729P&148 &-15 &-0.56 &-0.25 &            &            &            &            &            &          8 \\
 HD 114762P&165 &-14 &-0.75 &-0.70 &       0.43 &       0.32 &       0.20 &       0.24 &       0.30 &          5 \\
 HD 144579 &176 &-18 &-0.63 &-0.70 &       0.37 &       0.28 &            &            &            &          6 \\
 HD 218209 &188 &-21 &-0.54 &-0.43 &       0.19 &       0.18 &            &            &            &          6 \\
 HD 221830 &122 &-13 &-0.51 &-0.45 &       0.32 &       0.22 &       0.19 &       0.28 &       0.25 &          1 \\
 HD 230409 &118 & -6 &-0.84 &-1.00 &            &            &            &            &            &          4  \\
\hline
\end{tabular}  
\end{table*}

\begin{figure}
\resizebox{8cm}{10cm}{\includegraphics*{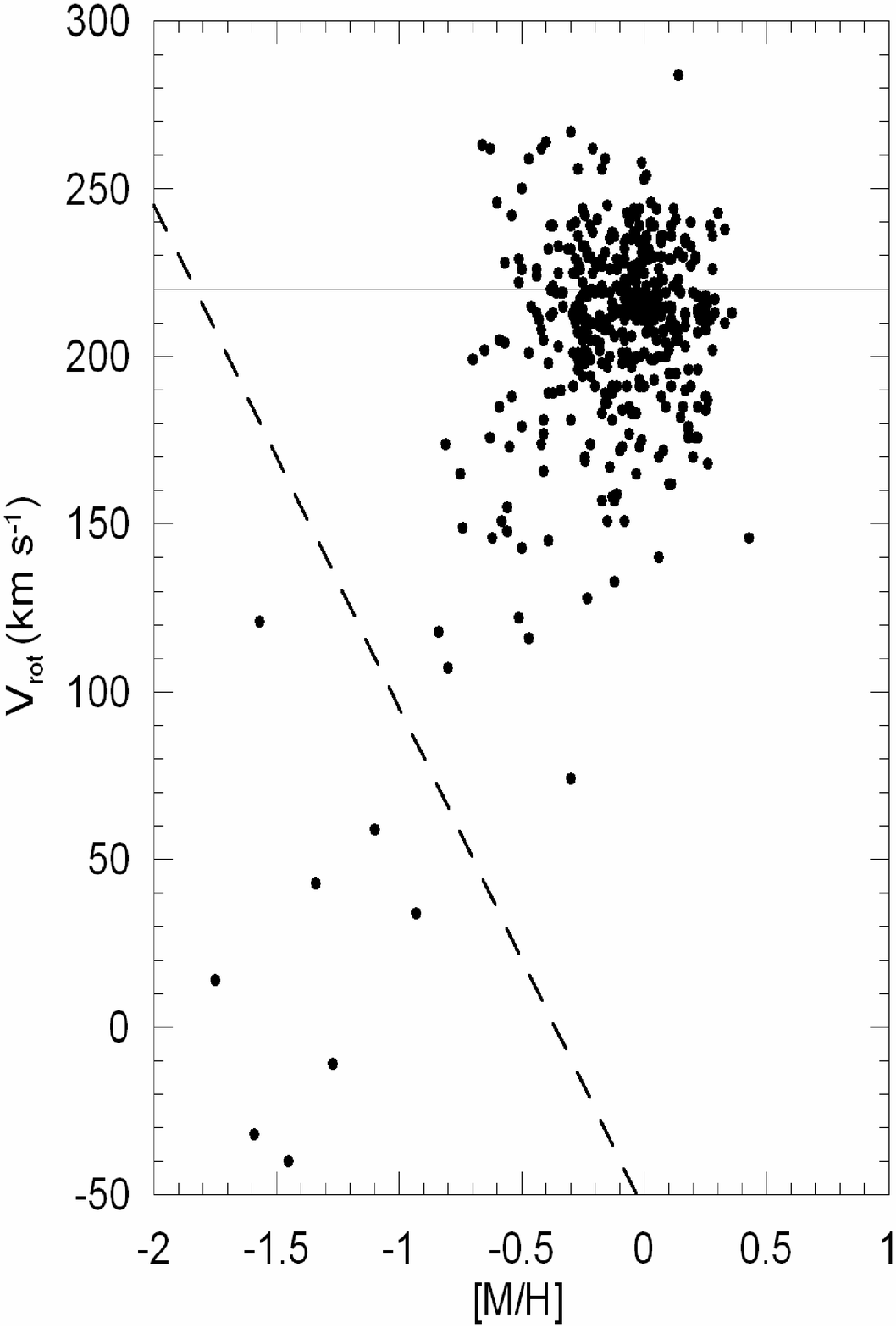}}
\caption{The $V_{\rm rot}$ versus [M/H] plot for our stars.
Here, $V_{\rm rot} = V' + $220 km $s^{-1}$.
The horizontal line corresponds to $V_{\rm rot}=$~220 km s$^{-1}$ of the LSR.
The dashed diagonal line is that separation criterion of Nissen \& Schuster (1991)
to separate halo from `high-velocity disk' stars.}
\end{figure}

In Fig.~6 the $V_{\rm rot}$, [M/H]  diagram for our sample stars is plotted.
Two components are clearly noted: the halo component,
$V_{\rm rot} \la $ 100 km s$^{-1}$  and $[M/H] \la -0.9$ dex; the other disk-like,
centered at $V_{\rm rot} \sim $~215 km s$^{-1}$ and [M/H] $\sim 0.0$ dex. The rotation
velocity of the LSR about the Galactic center is taken here to be
220 km s$^{-1}$ so that $V_{\rm rot} = V' + $~220 km s$^{-1}$ is
the rest-frame rotation velocity of a given star (Kerr \& LyndenBell 1986).
More modern values for the circular speed of the Milky Way at the solar circle
are somewhat larger, such as $\Omega_{\rm 0} = 234 \pm 13$ km s$^{-1}$ given
by Fukugita \& Peebles (2004).
Here, $V'$ is the velocity with respect to the LSR. It can be seen from Fig.~6 that
the majority of the stars have $V_{\rm rot} \sim $~200 km s$^{-1}$ and
$[M/H] \geq -0.5$ dex. These stars are thin-disk stars.
Even though there is scarcity of stars
with $[M/H] < -0.5$ dex and $V_{\rm rot}<$~150 km s$^{-1}$, $V_{\rm rot}$
increases approximately linearly with [M/H]  between $-1.0<[M/H]<-0.5$ dex.
For the $[M/H] < -1.0$ dex, it appears that there is no correlation
between [M/H] and $V_{\rm rot}$, but there are too few stars for any firm
conclusion.  

This $V_{\rm rot}$, [M/H] diagram can be used to separate out the
different stellar populations, as discussed in Nissen \& Schuster (1991) and in
Schuster et al. \ (1993). In the former reference, a diagonal cut connecting
([M/H], $V_{\rm rot}) = (-0.3,0$ km s$^{-1}$) and $(-1.5,175$ km s$^{-1}$) was used to
separate halo stars from `high-velocity disk' stars. In the latter reference,
the parameter ``$X$'' was defined, which is a linear combination of $V_{\rm rot}$ and [M/H],
and is used to make other diagonal cuts which isolate more cleanly the thick-disk
stars.  Our most recent work (Schuster et al. \ 2005) indicates that the range
$-21 \leq X \leq -6$ gives a fairly clean thick-disk sample, with only small contamination
by the halo and old thin disk.  In Schuster et al. \ (1993) the cut 
$-21 \leq X \leq -18$ was used to define an even cleaner thick-disk sample, but here
too few stars are found in this reduced $X$ interval.

The $X$ criterion provides a cleaner separation of the thin- and thick-disk components
even though they overlap considerably in both $V_{\rm rot}$ and $[Fe/H]$.  Using the 
definition of $X$ given by Schuster et al. \ (1993), 
combining $V_{\rm rot}$ and [Fe/H], the $X$ values of the stars have been calculated.  
The $X$ distribution of the stars is given in Fig.~7. 

Comparing our $X$ distribution with the one plotted by Schuster et al. \ (1993), 
constructed from a larger data base, $X$ cut-offs for the 
three stellar components are fixed for Fig.~7.  The number of stars in the thick-disk sample,
defined by the range, $-21 \leq X \leq -6$,  is 22. These limits for the thick disk have 
been derived from the work of Schuster et al. \ (1993) and from 
a project that is in process (Schuster et al. \ 2005). The thin disk dominates for 
$X \leq -33$, and the halo with a small fraction of the total extends for $X \geq -5$. 
These fixed cut-offs of three stellar populations are indicated as dashed vertical lines on 
Fig.~7.

The vertical velocity dispersion, $\sigma_{\rm W} = 32 \pm 5$ km s$^{-1}$ 
(error calculated according to Appendix E of Taylor \  1997), 
is estimated for the range $-21 \leq X \leq -6$, where the thick disk dominates. 
This value is less than the $\sigma_{\rm W} = 46$ km s$^{-1}$ of Schuster et al. \ (1993), 
but is in agreement with the range of 30--37 km s$^{-1}$ given by Norris (1987),
Croswell et al.~(1991), and Carney, Latham \& Laird (1989).  Other values, 
$<V_{\rm rot}> = 154 \pm 6$ km s$^{-1}$ and $<[M/H]> = -0.55 \pm 0.03$ dex (mean errors) for 
this thick-disk sample are in concordance with the mean values given in the literature, such 
as in Schuster et al. \ (1993).

\begin{figure}
\resizebox{8cm}{8cm}{\includegraphics*{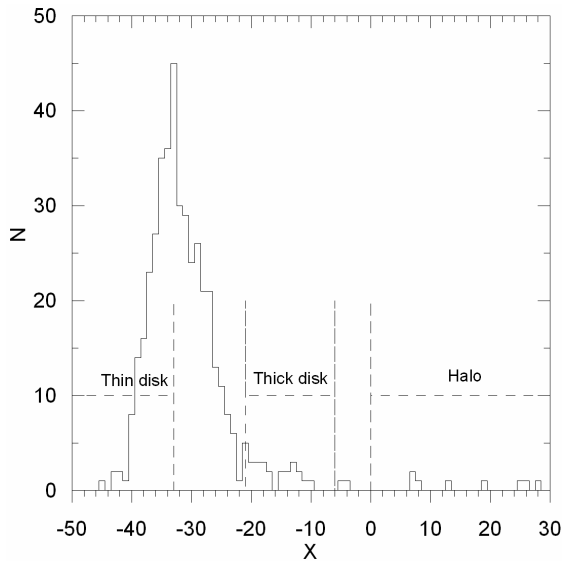}}
\caption{Distribution of the $X$ criteria, which combines the kinematic and
metallicity information of our sample. Dashed vertical lines have been drawn to illustrate 
the cut-offs of the thin and thick disks, and the halo.}
\end{figure}

Our thick-disk sample defined by the range, $-21 \leq X \leq -6$, is given in Table 3
together with the kinematic and abundance data. The $\alpha$-element and 
spectroscopic [Fe/H] abundances of some stars have been found in the literature. 
It can be seen from Table 3
that spectroscopic [Fe/H] abundances are in good agreement with the photometric ones.
The mean $<[\alpha/Fe]>$ values of seven thick-disk stars, except HD 16623 are constant at
a level of 0.27 dex, which are overabundant relative to thin disk stars, 
over the spectroscopic metallicity range from -0.45 to -0.84 dex (See Fig.~1.4 in Nissen 2004,
Fig.~4 in Nissen \& Schuster 1997, Fig.~13 in Bensby et al.\ 2003, and
Fig.~8 in Bensby et al.\ 2005).

[Mg/Fe] abundances of stars with thick-disk kinematics in Table 3, which fall in the range 
$+0.19 \leq [Mg/Fe] \leq+0.43$ dex are similar to those of the thick disk found in the literature.
The trend of $\alpha$-capture elements like Mg, Si, Ca, and Ti with 
photometric [M/H] (or spectroscopic [Fe/H]) abundances of thick-disk stars in 
Table 3 agrees well with the findings of Nissen \& Schuster (1997) and Bensby et al.~(2005).
Also, the $\alpha$-element abundances of thick-disk stars, which are available in Table 3,
support the classification procedure based on $X$ criteria defined by 
Schuster et al. \ (1993).

The only expection might be HD 16623; its $\alpha/Fe$ is low by a factor of 1.5 relative to the
seven other thick-disk stars in Table 3. The kinematics and photometric (and spectroscopic) [M/H] 
abundance of HD 16623 indicate that this star is a a member of the thick disk. The mean $[\alpha/Fe]$ 
of HD 16623 is low due to its [Ca/Fe] and [Ti/Fe] abundances, which are more similar to that of
the thin disk, whereas its [Mg/Fe] abundance is higher and more similar to that of the thick disk.
This star is perhaps transitional between the thick and thin disks, with a more mixed chemical and
dynamical history.

Another star which stands out in Table 3, and also in Fig.~6, is HD 104556.  Its $X$, $[M/H]_{phot}$,
and W' (= +15 km s$^{-1}$, from Table~2) values would all indicate a (thick) disk star while its
$V_{\rm rot}$ value (+74 km s$^{-1}$) is more indicative of the halo.  For example, from Table~5 of
Schuster et al. \ (1993) the $V_{\rm rot}$ value of HD 104556 is
$2.1\sigma$ removed from the mean value of the thick disk, but only $0.5\sigma$ from that of the halo. 
Other studies have also given compositions for HD 104556 which are more thick-disk-like than halo-like,
such as Friel (1987), who gives $[Fe/H] = -0.64$ dex, and Eggen (1997), who gives $[Fe/H]=-0.55$ dex.
These stars point out the risks in using only a single parameter, such as $X$, [Fe/H], or $[\alpha/Fe]$,
to derive the population type of an individual star, and also show that some stars do not fit
cleanly into any one population type, according to all criteria.

\subsection{Ages and an age-metallicity relation}

Fig.~8 shows the position of all the thin-disk stars in our sample, with a mean 
metal abundance of $<[M/H]>=-0.04$ dex for the metallicity range $-0.43\leq [M/H] \leq +0.35$ dex,
as compared with the isochrones of VandenBerg (2003) as transformed to the $uvby$
photometry by Clem et al.~(2004). It is seen that the majority of
the stars are distributed inside the isochrones.
Possible systematic shifts between the isochrones and the positions of the redder 
main-sequence stars have been examined, and no need for systematic corrections was found.
It is also conspicuous that a few stars with the values of $(b-y)_{o}\ga 0.44$ and
$M_{\rm V}\ga 4.5$ lie off the isochrone grid, i.e. ages $\ga 16$ Gyr. These may be
unidentified binary stars.

\begin{figure}
\resizebox{8cm}{8cm}{\includegraphics*{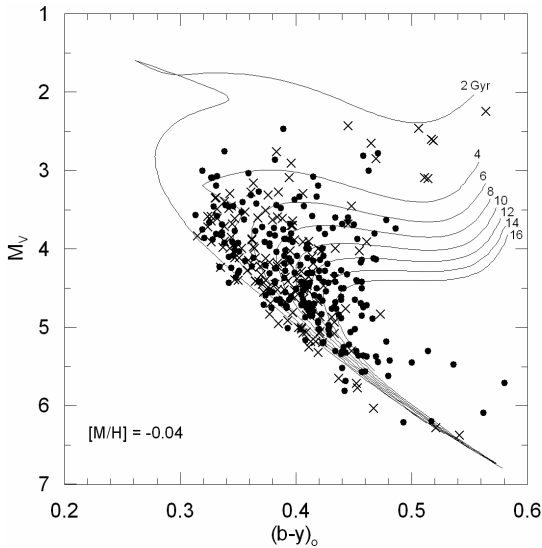}}
\caption{Solid lines represent the isochrones (from 2 Gyr to 16 Gyr in steps of 2 Gyr)
of VandenBerg (2003) as transformed to the $uvby$ 
system by the colour-temperature relations of Clem et al.~(2004); those isochrones 
corresponding to $[M/H]=-0.04$ dex have been plotted, which is the mean metal abundance 
of the thin-disk stars ($-0.43\leq [M/H] \leq +0.35$ dex) for $X \leq -33$ in our sample, the
definite thin-disk stars, shown as dots in this figure. Cross symbols show the 
probable thin-disk stars with $-33 < X < -21$.}
\end{figure}

Isochrone ages are determined by placing the stars in observational HR diagrams
(such as Fig.~8) using the observed $M_{\rm V}$, $(b-y)_{o}$ values for each star, and
then reading off the age of the star by interpolating between the theoretically computed
isochrones.  The isochrones of VandenBerg (2003) as transformed to the $uvby$ system 
by the colour-temperature relations of Clem et al.~(2004) have been chosen for the age 
determination of the thin-disk stars. These $uvby$ isochrones cover the range
$-0.60<[M/H]<+0.50$ dex and ages from 1 to 20 Gyr in 1 Gyr steps.  Whereas, for the age 
determination of the thick-disk and halo stars, which are separated according to the $X$
criterion, the set of isochrones from Bergbusch \& VandenBerg (2001), as transformed by
the same colour-temperature relations, have been used. This latter set of isochrones cover
$-2.31<[M/H]<-0.30$ dex and ages from 6--20 Gyr in 1 Gyr steps.  The ages of the 22
thick-disk stars have been interpolated in the $M_{\rm V}$, $(b-y)_{o}$ diagram using a
constant $[\alpha/Fe]= +0.30$ dex and interpolated for the metal abundances
of $[M/H]=-0.30, -0.40, -0.52, -0.61, -0.70, -0.83, -1.01$ dex. At sub-solar 
[M/H] we assumed $[\alpha/Fe]= +0.30$ dex for thick-disk stars with 
$[M/H]\la -0.8$ dex, taking into consideration Fig.~3 of Wheeler, Sneden \& Truran (1989).

\begin{figure}
\resizebox{8cm}{12cm}{\includegraphics*{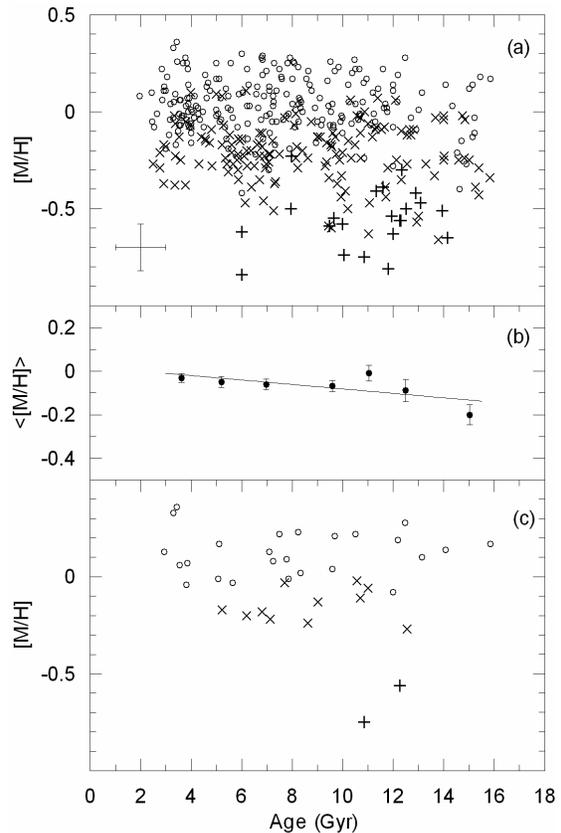}}
\caption{(a) Age-metallicity relation for our stellar sample. The circles show the stars with
$X \leq -33$, which are the definite thin-disk stars, Cross signs indicate stars with
$-33<X<-21$, which are probable thin-disk stars but with some small probability of 
being thick disk, and the plus signs show the probable thick-disk stars, respectively.
A typical average error bar, referring to the uncertainties of 1 Gyr in age and $0.12-0.14$ dex 
in Schuster \& Nissen (1989)'s metallicity calibration, is indicated.
(b) Average metallicity $<[M/H]>$ in age bins. The straight line is a least-squares fit having 
a slope of $-0.01 \pm 0.005$ dex Gyr$^{-1}$. Error bars are the standard deviation of the mean of the 
values in each bin. (c) Age-metallicity relation for 38 main-sequence stars which have
extra-solar planets according to the catalogues of Schneider (2005) and Mayor et al.~(2005)}
\end{figure}

In order to estimate the errors for the age interpolations, the uncertainties
($\sigma_{\rm V}$, and $\sigma _{b-y}$) in $V$, and $(b-y)$ have been collected
from the web site of the General Catalogue of Photometric Data of 
Hauck \& Mermilliod (1998), and  from the $Hipparcos$ and $Tycho$ catalogues (ESA\ 1997),
the uncertainties ($\sigma_{\pi}/\pi$) in the parallaxes ($\pi$) of our sample of
stars given in Table 2.
The derived error equation for the Pogson relation given in Section 4.2,
$\sigma _{M_{\rm V}}=\sigma_{\rm V} + 2.17(\sigma_{\pi}/\pi)$, allows us to estimate
$\sigma _{M_{\rm V}}$ from the uncertainties in $V$ and $\pi$ of individual stars
in our sample. The resulting average uncertainties in $M_{\rm V}$ and $(b-y)$ are
0.081 and 0.003 mag, respectively. For stars lying along the turn-offs of the isochrones,
these average uncertainties in $M_{\rm V}$ and $(b-y)$ result in age errors of $\sim$ 1 Gyr when
applied to the difference between two isochrones in the $M_{\rm V}$, $(b-y)_{o}$ 
diagram.  For sub-giants above the turn-offs, the errors are somewhat larger,
$1.0-1.5$ Gyr.  Below the turn-offs the age errors can grow indefinitely due to the
convergence of the isochrones.  We have measured ages to an error limit of about
$\pm 2.5$ Gyr.  Error bars corresponding to an uncertainties of 1 Gyr in age 
and $0.12-0.14$ dex in $[Fe/H]$, (Schuster \& Nissen 1989 and Section~4.3) are 
indicated in Fig.~9a.

In Fig.~9a the metallicity is plotted as a function of age with different symbols
for the definitive thin disk (circles, $X \leq-33$), probable thin disk 
(cross signs, $-33 < X < -21$) and probable thick disk (plus signs, $-21 \leq X \leq-6$). 
It can be seen from Fig.~9a that few stars have ages less than 3 Gyr; 
this is to be expected since the stars of this sample by Nidever et al.~(2002) 
contain mostly late F-, G-, and K-type stars. 
This derived age-metallicity relation agrees qualitatively with that
of Edvardsson et al.~(1993). 

A slight trend that metallicity decreases with increasing age is seen in Fig.~9b;  
the slope of a least squares fit to the $<[M/H]>$ versus age is $-0.01 \pm 0.005$ dex Gyr$^{-1}$.
Even though there is this overall trend of decreasing mean metallicity with increasing age,
both a larger scatter in [M/H] at a given age and the presence of old,
metal-rich stars make it difficult to decide whether or not an age-metallicity relation
really exists for the thin disk. So, it is difficult to draw any clear conclusion from 
Fig.~9a.

The age-metallicity relations derived from the larger samples of Feltzing et al.\ (2001) 
and Nordstr\"om et al.~(2004) apply mainly for stars less than 3 Gyr, whereas ages of our
stars are generally greater than this limit.  For ages larger than 3 Gyr the 
age-metallicity diagrams of Feltzing et al. ~(2001; their Fig.~10) and of
Nordstr\"om et al.~(2004; their Fig.~27) appear very similar to our Fig.~9a, almost no
correlation between age and metallicity.

In Fig.~9a the 22 thick-disk stars show a range of 7--8  Gyr in their ages but no obvious
correlation between the ages and metallicities, in contrast with the results of
Bensby, Feltzing \& Lundstr\"om (2004), who find a pronounced age-metallicity relation for the
thick disk.  However, Bensby et al.\ (2004) work with a thick-disk sample ten times 
larger than ours, separated from the thin disk more rigorously, and so their conclusions 
should be preferred.  We do agree with them that the thick-disk stars show a very 
significant range in ages, $\ga 5$ Gyr.

From Fig.~9a, it can be seen that there is lack of metal-poor young stars
in the left-bottom corner. However, there exists numerous old metal-rich stars
in the right-upper corner, which leads to the substantial scatter and the flattening of
the plot. The existence of the old metal-rich stars can also seen
in the works of Feltzing et al. \ (2001), Nordstr\"om et al.~(2004),
Rocha-Pinto et al.~(2000), and Chen et al.~(2003).

Fig.9c shows the same age-metallicity plot for the main sequence stars from
Nidever et al.~(2002) for which extra-solar planets have been detected (Schneider 2005;
Mayor et al.~2005).  Again, the relation between the metallicity and the age is very flat 
as in Fig.~9a, the main difference being that the large majority of these stars have high 
metallicities, $[M/H] \ga -0.2$ dex, as has been pointed out in many studies, such as 
Gonzalez (1997), Laughlin (2000), Santos, Israelian \& Mayor (2001), 
Murray \& Chaboyer (2002), and Reid (2002).  However, two of these stars (HD 114729 and 
HD 114762) have lower metallicities, $[M/H] \sim -0.6$ dex, and have been classified thick disk, 
with rather large ages, $\sim 12$ Gyr.  Interestingly, this entire sample of stars with 
extra-solar planetary systems has large ages and a large range in ages, 
$3 \la$ age $\la 13$ Gyr.

\begin{figure}
\resizebox{8cm}{13cm}{\includegraphics*{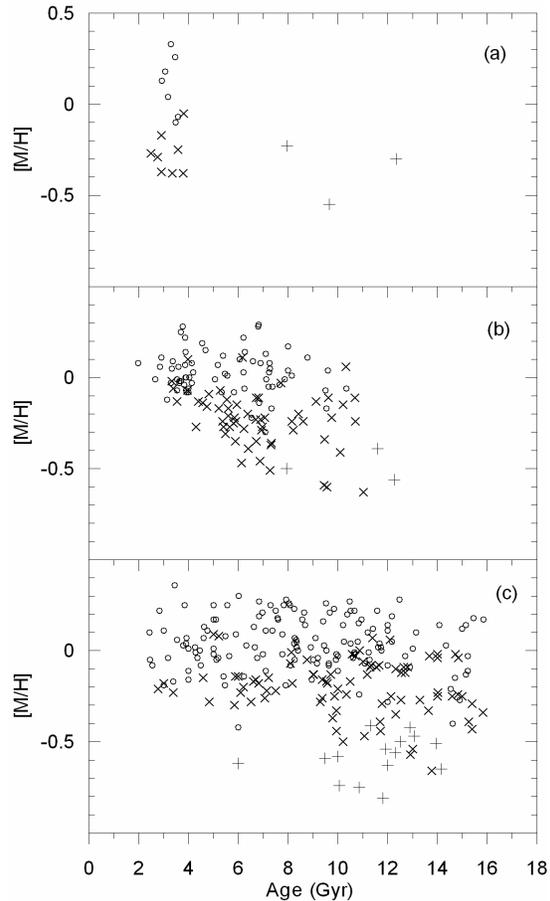}}
\caption{Age-metallicity plots of three sub-samples for the 
(a) stars with $M_{\rm V} \leq3$, (b) stars with $3<M_{\rm V}\leq4$, 
(c) stars with  $4<M_{\rm V}<6.3$. The symbols are the same as in Fig.~9a.}
\end{figure}

In Fig.10 we plot the ages and metallicities for our total sample, for three bins 
of $M_{\rm V}$.  A similar pattern as in Fig.14 of Feltzing et al.~(2001) appears. As 
we allow stars with faint absolute magnitudes to enter our samples,  the upper right-hand
corner of the age-metallicity plot is progressively filled in. In panel (c) of Fig.~10,
stars with $[M/H]>-0.1$ dex and ages larger than 10 Gyr become numerous in the upper
right-hand  corner. From panel (b) of Fig.~10, one might conclude a fairly obvious and
pronounced age-metallicity relation, but with the stars of panel (c) this has mostly
disappeared.  Fig.~10c reveals that old, metal-rich stars are dominant at the
faint absolute magnitudes.

Chen et al.~(2003) investigate the nature of these old metal-rich stars in the solar
neighbourhood and conclude that most seem to have originated from the inner thin disk.
They note that HD 190360 is an exception in their sample and suggest that it is a
metal-rich thick-disk star. A 15.8 Gyr age can be attributed to this star since its
position in the $M_{\rm V}$, $(b-y)_{o}$ diagram of Fig.~8 places it slightly to the
left of the 16 Gyr isochrone.  Our photometric metal abundance ($[M/H]=+ 0.17$ dex) and a
15.8 Gyr age show that this star is old and metal-rich, in accordance with the suggestion of
Chen et al.~(2003).  For this star, our kinematics ($V_{\rm rot} = $190 km s$^{-1}$, 
$W' = -56$ km s$^{-1}$) and photometric abundance ($[M/H]=+ 0.17$ dex) are almost identical 
to the ones of Chen et al.~(2003).   Whereas our $X=-35$ value indicates that it is a
thin-disk star, the rather large absolute value for $W'$ would suggest thick disk, again
in agreement with Chen et al.~(2003).  HD 190360 is another star which does not fit cleanly
into all the criteria for a single stellar population, and shows again the risk in using 
only two variables, [M/H] and $V_{\rm rot}$, to define the population type of a single star.

The scatter observed in age-metallicity plot of Fig.~9a can be interpreted within the scope
of models of Galactic chemical evolution of Pilyugin \& Edmunds (1996) and
Raiteri, Villata \& Navarro (1996). These models consider the chemical enrichment with time,
together with the dynamical evolution.  The model of Pilyugin \& Edmunds (1996)
specifically takes into account both the self-enrichment of star forming regions with 
sequential star formation and the effect of irregular rates of infall of un-processed 
matter onto the stellar disk.  These Galactic chemical-evolution models can
generate a very conspicuous scatter, such as that seen in the observational age-metallicity
plots, such as our Fig.~9a, Fig.~14a of Edvardsson et al.~(1993), Figs.~27--28 of
Nordstr\"om et al.~(2004), and Figs.~10--11 of Feltzing et al. \ (2001).  However, for
the models of Pilyugin \& Edmunds (1996) episodic gas infall onto the disk is required
to produce a scatter in metallicity ([Fe/H]) as large as that observed; self-enrichment
processes alone will not do it.  The N-body/hydrodynamical simulations of the chemical
evolution by Raiteri et al. \ (1996) can produce a scatter in [Fe/H] even larger than
that observed (see their Fig.~9) depending upon the volume in the Galaxy over which
the results have been considered.

Our sample based on radial velocities of Nidever et al.\ (2002) are not well
populated at all ages, but contains late-F, G and K stars with apparent magnitudes
fainter than $V = 8$ mag, as compared to the samples of Edvardsson et al.~(1993),
Feltzing et al. \ (2001) and Nordstr\"om et al.~(2004). This selection 
effect causes a bias in the sense that ages less than 3 Gyr are lacking, but the sample 
is without kinematic bias.  Moreover, the metallicity distribution of our sample is
representative of the older stellar populations of the solar neighbourhood, and shows
a small overall trend in the age-metallicity plot. The presence of old, metal-rich stars,
from the cooler program stars, shows that this sample has avoided one of the distortions
of the age-metallicity relation, such as that which affected the results of
Twarog (1980), as pointed out by Feltzing et al. \ (2001).  Also, the 
presence of a significant number of metal-rich and older stars emphasizes the usefulness of
the planetary-search list of Nidever et al.~(2002) since many studies, such as those of
Gonzalez (1997), Laughlin (2000), Santos et al. \ (2001),
Murray  \&  Chaboyer (2002), and Reid (2002), have shown that the parent stars
of most extra-solar planets are significantly metal-rich, as confirmed in our Fig.~9c.

\section{Conclusions}

Our main conclusions are as follows:

1) Accurate metal abundances can be derived from the $uvby-H\beta$ photometry,
using the combined calibrations as described above in Section~4.3.  In Fig.~4 a comparison 
between the photometric metallicities and modern spectroscopic ones for 71 stars shows
excellent agreement:  $[M/H]_{\rm spec}=0.99(\pm0.04)[M/H]_{\rm phot} - 0.01 (\pm0.02)$  for
the metallicity range $-2.0 <[M/H]<+0.5$ dex with a scatter of $\sigma_{[M/H]} = \pm 0.12$.
This error agrees well with estimates made by Feltzing et al. \ (2001),
$\pm0.10$--0.11.

2) The metallicity and $V_{\rm rot}$, [M/H] distributions of our sample show a mixture of all
the stellar populations in the solar neighbourhood; the thin disk dominates with a small but
significant contribution from the thick disk, plus a few halo stars.

3) Using the $X$ criterion of Schuster et al. \ (1993) has allowed
us to isolate a thick-disk sample of 22 stars with minimal contamination from the other stellar
populations.  A $\sigma_{\rm W} = 32 \pm 5$ km s$^{-1}$ velocity dispersion is estimated for
the range $-21\leq X \leq -6$ where these thick-disk stars dominate. This value is
within the range of 30--37 km s$^{-1}$ given by Norris (1987), Croswell et al.~(1991),
and Carney et al. \ (1989).  These 22 thick-disk stars also have
$<V_{\rm rot}> = 154 \pm 6$ km s$^{-1}$ and $<[M/H]> = -0.55 \pm 0.03$ dex.

4) $\alpha$-element abundances which are available for some thick-disk stars in our sample
show the typical $\alpha$-element signatures of the thick disk, supporting the 
classification procedure based on the $X$ criteria, which is a linear combination of
$V_{\rm rot}$ and [Fe/H] (Schuster et al. \ 1993).

5) There is a lack of stars with ages less than 3 Gyr in our age-metallicity diagram, due
to the selection criteria of the sample.  For ages greater than 3 Gyr, both a larger 
scatter in $[M/H]$ at a given age and the presence of old, metal-rich stars make it 
difficult to decide whether or not an age-metallicity relation really exists, very similar 
to the works of Edvardsson et al.~(1993), Feltzing et al. \ (2001) and 
Nordstr\"om et al.~(2004). Our age-metallicity plot (Fig.~9) shows a very shallow slope 
of $\Delta ([M/Fe])/\Delta$(age)$ = -0.01 \pm 0.005$ dex Gyr$^{-1}$.

6) Our Fig.~9a shows a significant age span for the thick disk with a range
in ages $\ga 5$ Gyr, in agreement with the work of Bensby et al. \ (2004).

7) The age-metallcity diagram for the stars with extra-solar planets shows a relation very
similar to that of the total sample, very flat with almost no variation in the mean
metallicity with age; the main difference is that most of these stars are metal-rich,
$[M/H] \ga -0.2$ dex with the exception of two (HD 114729 and HD 114762) which have 
$[M/H] \sim -0.6$ dex and have been classified as thick disk.  This subsample also shows a 
more or less uniform distribution over a wide range in ages, $3 \la$ age $\la 13$ Gyr, 
similar to the total sample.

8) A number of stars have been pointed out which do not fit well into any stellar-population
classification scheme, such as HD 16623, HD 104556, and HD 190360.  These stars exemplify the
risks in using only a single criterion, such as $X$, [Fe/H], $V_{\rm rot}$, [$\alpha$/Fe], or
$W'$, to classify the stellar-population-type of any individual star.  All such information
should be used, but even then, some stars cannot be classified unambiguously.

\section{Acknowledgments}

This research has been made possible by the use of the SIMBAD database, 
operated at the CDS, Strasbourg, France, and the web site of the General Catalogue of
Photometric Data, Geneva, Switzerland. We are indebted to D.A. Vandenberg
and J.L. Clem for kindly providing their isochrones. T. Bensby is thanked for providing 
his abundance data set. We also thank H. \c{C}akmak for preparing a visual basic program 
for the age determination. We would like to thank Bernard E. J. Pagel, the referee,
for his useful and constructive comments concerning the manuscript.
This work was supported by the Research Fund of the University of Istanbul,
Project number: BYP-402/26042004.

\bsp


\begin{thebibliography}{99}


\bibitem{}Allen C., Santillan A., 1991, Rev. Mex. Astron. Astrofis., 22, 255

\bibitem{}Arce H.G., Goodman A.A., 1999., ApJ, 512, L135

\bibitem{}Bensby T., Feltzing S.,  Lundstr\"om I., Ilyin I., 2005, A\&A, 433, 185

\bibitem{}Bensby T., Feltzing S., Lundstr\"om I., 2004, A\&A, 421, 969

\bibitem{}Bensby T., Feltzing S., Lundstr\"om I., 2003., A\&A, 410, 527

\bibitem{}Bergbusch P. A.,  VandenBerg D.A., 2001, ApJ, 556, 322

\bibitem{}Butler R.P., Vogt S.S., Marcy G.W., Fischer D.A., Henry G.W., \&
Apps K., 2000, ApJ, 545, 504

\bibitem{}Carney B., Latham D.W., Laird J.B., 1989, ApJ, 97, 423

\bibitem{}Cayrel de Strobel G., Soubiran C., Ralite N., 2001, A\&A, 373, 159

\bibitem{}Cenarro A.J., Cardiel N., Gorgas J., Peletier R.F., Vozdekis A.,
Prada F., 2001, MNRAS, 326, 959

\bibitem{}Chen Y. Q., Zhao G., Nissen P. E., Bai G. S., Qiu H. M., 2003, ApJ, 591, 925

\bibitem{}Chen Y. Q., Nissen P. E., Zhao G., Zhang H. W., \& Benoni T., 2000, A\&A, 
141, 491

\bibitem{}Clem J.L., VandenBerg D.A., Grundahl F., Bell R.A., 2004, AJ, 127, 1227

\bibitem{}Crawford D.L., 1975, PASP, 87, 481

\bibitem{}Croswell K., Latham D.W., Carney B., Schuster W., Aguilar L., 1991, AJ, 101, 2078

\bibitem{}Edvardsson B., Andersen J., Gustafsson B., Lambert D.L., Nissen P.E.,
 Tomkin J., 1993, A\&A, 275, 101

\bibitem{}Eggen O.J., 1997, AJ, 114, 825

\bibitem{}ESA 1997, {\it The Hipparcos and Tycho Catalogues}, ESA SP-1200, Noordwijk

\bibitem{}Feltzing S., Holmberg J., Hurley J.R.,  2001, A\&A, 377, 911

\bibitem{}Friel E.D., 1987, AJ, 93, 1388

\bibitem{}Fukugita M.,  Peebles P.J.E., 2004, ApJ, 616, 643

\bibitem{}Fulbright J., 2000, AJ, 120, 1841

\bibitem{}Gim\'enez A., 2000, A\&A,356, 213

\bibitem{}Gonzalez G., 1997, MNRAS, 285, 403.

\bibitem{}Hauck B., Mermilliod M., 1998, A\&AS, 129, 431

\bibitem{}Henry G.W., Fekel F.C., Henry S.M.,  Hall, D.S., 2000, ApJS, 130, 201

\bibitem{}Hog E., Fabricius C., Makarov V. V., Urban S., Corbin T., 
Wycoff G., Bastian U., Schwekendiek P., Wicenec A., 2000, A\&A, 355L, 27

\bibitem{}Johnson D.R.H.,  Soderblom D.R., 1987, AJ, 93, 864

\bibitem{}Kerr F.J.,  Lynden-Bell D., 1986, MNRAS, 221, 1023

\bibitem{}Kharchenko N.V., 2001, Kinematika i Fizika Nebesnykh Tel, 17, 409

\bibitem{}Laughlin G., 2000, ApJ, 545, 1064

\bibitem{}Lutz T.E.,  Kelker D.H., 1973, PASP, 85, 573

\bibitem{}Martell S.,  Laughlin G., 2002, ApJ, 577, 45

\bibitem{}Mayor M., Naef D., Pepe F., Queloz D., Santos N.C., Udry S., 2005,
{\it The Geneva Extrasolar Planet Search Programmes}, Universit\'e de Gen\'eve,\\
http://www.unige.ch/sciences/astro/fr/Recherches.

\bibitem{}Mishenina T.V., Soubiran C., Kovtyukh V.V., Korotin S.A., 2004, A\&A, 418, 551

\bibitem{}Murray N., Chaboyer B., 2002, ApJ, 566, 442

\bibitem{}Nidever D.L., Marcy G.W., Butler R.P., Fischer D.A., Vogt S. S., 2002, 
ApJS, 141, 503

\bibitem{}Nissen P.E., 2004, in {\it Origin and Evolution of the elements from the Carnegie 
observatories centennial symposia}, eds. A. McWilliam,  \& M. Raucherro,
Cambrdige Univ. Press., Cambridge, Vol.4, 156

\bibitem{}Nissen P.E., 1994, in {\it Stars, Gas, and Dust in the Galaxy, Invited Reviews at a
Symposium in Honor of Eugenio E. Mendoza}, eds. A. Arellano Ferro \& M. Rosado,
Rev. Mex. Astr. Astrofis., 29, 129

\bibitem{}Nissen P.E., Schuster W.J., 1997, A\&A, 326, 751

\bibitem{}Nissen P.E., Schuster W.J., 1991, A\&A, 251, 457

\bibitem{}Nordstr\"om  B., Mayor M., Andersen, J., Holmberg, J., Pont, F.,
J{\o}rgensen, B.R., Olsen, E.H., Udry, S., \& Mowlavi, N., 2004, A\&A, 418, 989

\bibitem{}Norris J.E., 1987, ApJ, 314, L39

\bibitem{}Pilyugin L.S., Edmunds M.G., 1996, A\&A, 313, 792

\bibitem{}Pomp\'eia L., Barbuy B., Grenon M., 2003, ApJ, 592, 1173

\bibitem{}Raiteri C.M., Villata M., Navarro J.F., 1996, A\&A, 315, 105

\bibitem{}Reid I.N., 2002, PASP, 114, 306

\bibitem{}Rocha-Pinto H.J., Maciel W.J., Scalo J., Flynn C., 2000, A\&A, 358, 850

\bibitem{}Santos N.C., Israelian G., Mayor M., 2004, A\&A, 415, 1153

\bibitem{}Santos N.C., Israelian, G., \& Mayor, M., 2001, A\&A, 373, 1019

\bibitem{}Schlegel D.J., Finkbeiner D.P.,  Davis M., 1998, ApJ, 500, 525

\bibitem{}Schneider J., 2005, {\it Extra-solar Planets Catalog}.  Observatoire de Paris,
http://www.obspm.fr/encycl/cat1.html

\bibitem{}Schuster W.J., Moitinho A., Parrao A., M\'arquez, A., Covarrubias E., 2005,
A\&A, in process

\bibitem{}Schuster W.J., Beers T.C., Michel R., Nissen P.E., Garc\'{\i}a, G., 2004,
A\&A, 422, 527

\bibitem{}Schuster W.J., Nissen P.E., Parrao L., Beers T.C., Overgaard L.P.,
1996, A\&AS, 117, 317

\bibitem{}Schuster W.J., Parrao L., Contreras-Mart\'{\i}nez M.E., 1993, A\&AS, 97, 951

\bibitem{}Schuster W.J.,  Nissen P.E., 1989, A\&A, 221, 65

\bibitem{}Taylor J.R., 1997, {\it An Introduction to Error Analysis, The Study of
Uncertainties in Physical Measurements}.  University Science Books, Sausalito

\bibitem{}Twarog B.A., Anthony-Twarog B.J., Tanner D., 2002, AJ, 123, 2715

\bibitem{}Twarog B.A., 1980, ApJ, 242, 242

\bibitem{}VandenBerg D.A., 2003, private communication

\bibitem{}VandenBerg D.A., Clem J.L., 2003, AJ, 126, 778

\bibitem{}Vogt S.S., Marcy G.W., Butler R.P., Apps K., 2000, ApJ, 536, 902

\bibitem{}Wheeler J.C., Sneden C., Truran J.J.W., 1989, ARA\&A, 27, 279


\end{thebibliography}
\end{document}